## Biophysical Considerations for Rational Antibody and ADC Design

Alberto Ocana[1,2,3,4]* Jorge R. Espinosa[3,4,5,6]*

[1]Experimental Therapeutics Unit, Department of Medical Oncology, Hospital Clínico Universitario San Carlos, Instituto de Investigación Sanitaria San Carlos (IdISSC), Madrid, Spain and CIBERONC, Madrid, Spain. alberto.ocana@salud.madrid.org
[2] Cátedra START-INTHEOS-CEU. Facultad de Medicina, CEU San Pablo, Madrid. alberto.ocana@startmadrid.com
[3] Instituto Pluridisciplinar, Universidad Complutense de Madrid, Madrid, Spain
[4] C-Therapeutics, Houston, USA.
[5] Universidad Complutense de Madrid, Department of Physical Chemistry, Madrid, Spain. jorgerene@ucm.es
[6] University of Cambridge, Yusuf Hamied Department of Chemistry, Cambridge, UK

*Corresponding authors.

### Abstract

Antibody-based therapeutics—including antibody–drug conjugates (ADCs), bispecific antibodies, and novel formats—are reshaping oncology, yet key determinants of efficacy, safety, and manufacturability frequently emerge after conjugation and formulation. We argue that computational biophysics provides an underexploited framework to address this gap by connecting molecular interactions to biological outcomes. We highlight how molecular dynamics, coarse-grained simulations, and free energy calculations reveal how conjugation site, linker chemistry, and drug–antibody ratio reshape conformational landscapes. We emphasize structural coupling between antibody, linker, and payload, with implications for antigen binding, internalization, and developability. We propose that integrating physics-based modeling into development pipelines—alongside experimental validation—can reduce empirical iteration and de-risk translation. As force fields, and hybrid physics–machine-learning methods improve, this field is poised to become a central driver of next-generation ADC design.

### Significance statement

Antibody-based cancer therapies, including antibody–drug conjugates, frequently fail late in development due to properties that emerge only after chemical modification and formulation. We argue that computational biophysics—encompassing molecular dynamics, coarse-grained simulations, and free-energy calculations—provides an underexploited framework to predict these properties early, connecting molecular interactions to biological outcomes. Integrating physics-based modeling alongside experimental validation can reduce empirical iteration, define rational design windows, and accelerate the translation of next-generation antibody therapeutics to the clinic.

## Beyond static models: computational biophysics as a rational framework for antibody and ADC design

The rapid expansion of antibody-based cancer therapeutics—including antibody–drug conjugates (ADCs), bispecific and multispecific antibodies, and chemically engineered formats—has exposed fundamental limitations in current antibody design paradigms [1,2]. Although advances in sequence engineering and structure prediction have transformed early discovery, many determinants of clinical efficacy and failure emerge only after chemical modification, formulation, and multivalent self-assembly [1,2]. These determinants are governed by physical interactions that are poorly described by static or sequence-centric approaches, contributing to late-stage attrition in oncology drug development [3].

Current approaches optimize antibodies and ADCs largely as modular constructs, refining sequence, conjugation site, linker, and payload in isolation. Yet properties critical to clinical outcome—including aggregation propensity, receptor internalization efficiency, intracellular drug release, and manufacturability—arise from the dynamic, coupled interplay between all these components and their biological environment. Neither sequence engineering nor static structure prediction can fully capture these emergent, physically driven behaviours, which is why many design failures are only uncovered late in development, at significant cost.

Unlike conventional approaches, computational biophysics treats antibodies and ADCs as dynamic, ensemble-based systems rather than rigid scaffolds. Atomistic molecular dynamics, coarse-grained simulations, and free-energy calculations can directly connect molecular-level interactions—electrostatics, hydrophobicity, steric constraints—to higher-order organisation and biological outcomes such as antigen binding, internalization, and intracellular payload release. Critically, these methods can evaluate how conjugation site selection, linker chemistry, and drug–antibody ratio (DAR) collectively reshape the conformational landscape of the molecule, rather than treating each variable independently.

In this Perspective, we argue that integrating physics-based modelling into antibody and ADC development pipelines—alongside experimental cross-validation—can reduce empirical iteration, define viable design windows, and de-risk translation. We discuss how these methods apply across key stages: conjugation-site selection, linker and payload optimisation, multivalency tuning, and manufacturability assessment[4]. We contend that computational biophysics is not merely a supporting interpretive tool, but is positioned to become a central, decision-level driver of next-generation antibody and ADC design.

## Use of Computational Biophysics for Future-Generation Antibodies

### *All-atom molecular dynamics (MD) simulations*

At the molecular level, atomistic computational methods provide microscopic insight into the interactions that define antibody structure and functionality. All-atom molecular dynamics (MD) simulations explicitly model atomistic intermolecular forces between biomolecules (e.g., proteins, nucleic acids or lipids), water and ions, enabling the analysis of electrostatic effects, hydrogen bonding, hydrophobic contributions, and steric constraints [5,6]. Atomistic simulations can be particularly useful for evaluating payload accessibility and linker exposure, which are critical determinants of ADC efficacy [4,7,8]. For ADCs employing cleavable linkers, simulations can also estimate solvent and protein accessibility of enzymatic cleavage sites [8], informing whether intracellular proteases are likely to access and process the linker efficiently, as later described in the article. Accurate atomistic biophysical modeling crucially depends on the use of rigorously validated force fields, including CHARMM36m [9], OPLS-AA [10], and AMBER ff19SB [11] for protein-protein interactions represented in explicit solvent and ionic environments, together with CGenFF [12], OPLS-derived parameter sets [13] and GAFF [14] for consistent treatment of linker and payload general interactions. Force fields are the mathematical rules that define how atoms interact in a simulation — essentially the physical laws of the virtual molecular world. Using rigorously validated force fields, tailored separately for the protein, linker, and payload components, is essential to ensure that the simulation faithfully reproduces real chemical behaviour and generates trustworthy predictions. Simulations using these force fields can provide the rational basis for selecting and optimizing conjugation sites, linker architectures, and DARs that balance systemic stability with effective intracellular drug release, an issue which has been shown experimentally to strongly influence therapeutic index [15,16]. The therapeutic index reflects the balance between a drug's efficacy and its toxicity — a wider index means a drug can effectively kill tumour cells while causing minimal harm to healthy tissue. Getting conjugation sites, linker design, and DAR right is therefore not merely a technical optimisation: it directly determines whether an ADC is safe and effective enough to benefit patients in the clinic. In **Figure 1A**, we sketch the different variable and constant domains of an antibody—including both light and heavy chains—and how structural sequence predictors such as AlphaFold [17] or ColabFold [18] can generate an initial configuration suitable for performing an atomistic simulation in well-optimized and parallelizable open-access MD packages such as GROMACS [19], LAMMPS [20] or OpenMM [21]. We note that although structural prediction models such as AlphaFold or ColabFold can indeed generate full antibody structures, predictions for Fc-domain orientations, structures and flexible interdomain arrangements may be less reliable in some cases. Consequently, several studies have focused primarily on Fab domains to investigate antigen recognition and binding interaction profiles with higher confidence[22] . Regarding linker regions and intrinsically disordered segments, AI-based structural predictors generally identify these regions only in probabilistic terms—as ordered or disordered with a certain degree of confidence—without fully capturing their dynamic conformational ensembles[23] . In contrast, MD simulations enable explicit sampling of intrinsically disordered regions and flexible linkers, allowing exploration of their accessible conformational space and transient intermolecular interactions[24] Therefore, once an initial antibody or ADC structure is generated using AI-driven predictors such as AlphaFold, subsequent all-atom or high-resolution coarse-grained MD simulations are highly valuable to relax the structure, refine intra and intermolecular interactions, and characterize the ensemble behavior of flexible vs. globular domains.

In **Figure 1B** (orange panel; all-atom), we show some relevant properties which can be accessed through extensive atomistic simulation trajectories, such as the conformational dynamics of a given protein domain [25], the impact of single-point sequence mutations on the protein structural dynamics [26] or the kinetic dissociation constant between different protein domains [27], which can be extremely useful to predict antigen-antibody binding affinities. Importantly, despite their substantial computational cost, atomistic simulations uniquely provide access to realistic conformational ensembles and interaction networks, capturing the intrinsic flexibility of antibody protein domains and conjugated warheads beyond what can be inferred from static structural models [28,29]. Moreover, in the context of pH-dependent binding behaviour–where several therapeutic antibodies such as anti-PCSK9, anti-IL6R, and FcRn-interacting antibodies[30] [XX] display antigen binding affinity variations between physiological and endosomal pH conditions due to protonation-state changes in histidine–MD simulations at constant pH (in which the protonation state depends on the pH[31] are particularly valuable. All-atom simulations can explicitly model protonation equilibria together with structural fluctuations, conformational dynamics and electrostatic reorganization as a function of the precise pH[32] In contrast, current AI-based structural prediction methods largely rely on fixed protonation states and static conformational representations, limiting their ability to capture pH-coupled energetic landscapes, endosomal dissociation mechanisms, or protonation-driven binding switches. Therefore, pH-responsive antibodies represent a clear example where physics-based simulations at all-atom level provide mechanistic insight beyond the current scope of purely AI-driven prediction frameworks.

*Coarse-grained (CG) and multiscale biophysical models*

While all-atom MD simulations provide atomistic detail and a realistic representation of the intermolecular forces governing protein-protein, protein-linker or protein-payload interactions [33], coarse-grained (CG) and multiscale biophysical models address their intrinsic sampling limitations by reducing molecular degrees of freedom while preserving key physicochemical interactions, thereby enabling exploration of longer timescales and larger system sizes relevant to antibody and ADC phase behavior (**Figure 2A**) [34]. Rather than tracking the position and movement of every individual atom, coarse-grained models group clusters of atoms together into single, larger interaction units. This is analogous to zooming out on a map: fine local detail is sacrificed in exchange for a much broader view of the landscape. The practical consequence is a dramatic reduction in computational cost, allowing simulations to cover timescales and system sizes that would be completely inaccessible to all-atom approaches—minutes or hours of molecular behaviour rather than nanoseconds, and systems containing entire antibody assemblies rather than isolated domains. Crucially, this simplification is not arbitrary: the grouped units are designed to preserve the key physical and chemical properties—such as charge, hydrophobicity, and shape—that drive biologically relevant behaviour, so that meaningful predictions about antibody stability, aggregation, and self-assembly can still be made. By representing groups of atoms as single interaction sites [35–37], CG approaches allow systematic interrogation of global antibody conformational ensembles, including Fab–Fc flexibility and long-range domain coupling, which are difficult to capture using atomistic simulations alone. Simplified antibody representations, such as the 12-bead model, have been successfully applied to investigate intermolecular self-association and stability of monoclonal antibodies under formulation-relevant conditions, providing mechanistic insight into aggregation propensity and viscosity at high concentrations [38,39]. In **Figure 1B** (green panel), we summarize some additional properties which can be accessed through CG simulations, such as antibody/ADC clustering, internalization propensity, or antigen-antibody binding affinity [40,41]

In the context of ADCs, CG simulations extend this framework by revealing how conjugation-induced perturbations—including increased DAR or redistribution of payloads across antibody domains—reshape the global conformational landscape rather than simply adding chemical functionality [4]. Such conformational ensemble redistribution can influence recognition-domain structural variations, modulate multivalent interactions, and higher-order self-assembly with direct consequences for solvation behavior, manufacturability, and *in vivo* performance—key determinants of clinical success in cancer therapeutics [42]. Recent advances in high-resolution CG force fields, including MARTINI based frameworks, have further enabled the characterization of protein–excipient interactions using Fab domains from therapeutic antibodies such as trastuzumab and omalizumab as representative model systems [34]. By bridging submolecular-level interactions and macromolecular ensemble phase behavior, CG simulations at different resolution levels provide a scalable route to de-risk ADC design prior to experimental validation, complementing atomistic predictions at substantially larger length scales and timescales (**Figure 1 and 2**).

A further advantage of biophysical modeling lies in its ability to evaluate binding affinity across multiple, coupled interfaces that must be simultaneously optimized in ADCs [43,44]. Effective ADCs must preserve high-affinity antigen binding while accommodating payload attachment, maintaining linker accessibility for productive release, and in some cases, retaining Fc-mediated immune functions [43]. Hybrid computational frameworks, such as the Site-Identification by Ligand Competitive Saturation (SILCS) method, enable systematic, quantitative interrogation of these competing requirements by combining interaction mapping with ensemble-based analysis [45,46]. By relying on precomputed free-energy landscapes rather than repeating full-system simulations, these approaches— which often integrate grand canonical Monte Carlo with classical MD simulations to generate pre-computed grid free energy maps—allow rapid comparison of ADC variants differing in conjugation site, linker chemistry, or DAR, while explicitly accounting for payload–antibody and payload–payload interactions [4,47,48]. Moreover, they can provide relatively quick predictions of antibody binding site recognition and protein-excipient interactions, which are highly valuable before further advancing experimental validation (**Figure 1B**; cyan panel). Unlike conventional docking approaches that typically rely on a single static structure[49,50] SILCS explicitly incorporates protein flexibility and solvent competition, enabling ensemble-based characterization of transient interaction hotspots and partially cryptic binding regions. This capability is especially relevant for antibodies and ADCs, where conformational heterogeneities, payload shielding, linker accessibility, and excipient interactions can strongly influence developability and therapeutic performance[4] . Moreover, beyond the examples provided in **Figure 1B,** in the context of ADCs, SILCS-Biologics[51] becomes useful for identifying potential payload interaction regions, assessing conjugation-site environments, estimating excipient interaction tendencies, and evaluating how local physicochemical environments may compete with antigen recognition.

Within this framework, simulations can identify payload binding modes that sterically or energetically outcompete with antigen recognition, as well as payload clustering that induces ADC aggregation, or limits solvent exposure and enzymatic processing—mechanisms consistent with experimental observations linking increased hydrophobicity and DAR to reduced efficacy [15,16]. Importantly, these effects are often energetically and structurally coupled, such that improving one property (e.g., payload exposure or avidity) can inadvertently compromise another (e.g., antigen binding affinity or developability). By integrating free energy simulations, interaction mapping, and ensemble-level descriptors, multiscale biophysical modeling provides a coherent strategy to navigate these trade-offs across different resolution and length scales, as well as define feasible design windows for rational ADC optimization. However, it is always fundamental to cross-validate computational predictions with experimental values to ensure accurate modeling performance. In **Figure 2B**, we recapitulate several properties which can be benchmarked both computationally—with models of different resolution-levels—and experimentally to guarantee realistic modeling predictions and improve model parameterization when needed. Properties amenable to cross validation using experimental studies, include aggregation propensity, binding affinity, hydrodynamic radius, and viscosity, among others, that can directly benchmark the outputs of each modeling tier. Therefore, computational biophysics should function as a complementary, hypothesis-generating tool within an integrated experimental-computational pipeline, rather than as a standalone predictive framework.

**Antibody affinity, multivalency and internalization**

Computational characterization through MD simulations reframe antibody therapeutics as dynamic, ensemble-based systems rather than static molecular entities [33,52]. Antibodies and ADCs populate rich distributions of molecular conformations whose relative populations are sensitive to sequence variation, chemical modification, and environmental context. Conjugation can stabilize rare conformations or suppress functionally relevant states, with direct implications for tumor targeting, intracellular trafficking, and cytotoxic efficacy [3]. Ensemble-based analyses therefore provide a more realistic and predictive description of antibody mesoscopic phase behavior in complex biological environments. One of the clearest examples of internalization-driven therapeutic behavior is trastuzumab, a monoclonal antibody targeting HER2-positive breast cancer. Trastuzumab binds HER2 with high affinity but induces relatively slow receptor internalization and substantial receptor recycling, contributing to sustained receptor blockade and antibody-dependent cellular cytotoxicity (ADCC) [53]. While this internalization profile is advantageous for antibody-only mechanisms of action, it limits intracellular drug delivery, illustrating how internalization requirements differ fundamentally between naked antibodies and ADCs. HER2-directed ADCs further illustrate how internalization can be engineered through molecular design. Trastuzumab emtansine (T-DM1) and trastuzumab deruxtecan (T-DXd) both rely on the trastuzumab antibody backbone, despite HER2 being a relatively slowly internalizing receptor [54].

In these cases, conjugation to cytotoxic payloads with optimized linkers and DAR enable sufficient intracellular accumulation for efficacy. In particular, trastuzumab deruxtecan combines a higher DAR with a membrane-permeable payload, allowing partial compensation for moderate internalization through bystander killing effects [55]. Similarly, multivalency plays a central role in dictating internalization behavior across other approved ADCs. Sacituzumab govitecan, which targets TROP2, leverages high target expression and multivalent engagement to promote receptor clustering and subsequent internalization [56]. The combination of moderate affinity, high receptor density, and flexible linker chemistry supports efficient uptake while avoiding excessive receptor downregulation. Computational biophysical evaluation through MD simulations—either at atomistic level or using high-resolution chemically-accurate CG models [35]—can help tune multivalency, and directly relate chemical sequence modifications across the ADC with their corresponding impact on intermolecular avidity and binding affinity. Excessive avidity or clustering can induce non-productive receptor aggregation, altered trafficking, or toxicity, whereas insufficient multivalent engagement may fail to trigger endocytosis altogether. Hydrophobic payloads, such as auristatins and maytansinoids, can promote nonspecific membrane interactions that enhance uptake but also increase aggregation risk and off-target toxicity [15,16]. Strategies employed in later-generation ADCs—including increased linker hydrophilicity, controlled DAR, and site-specific conjugation—reflect an implicit biophysical optimization of internalization versus systemic stability.

**Site specific conjugation**

Computational and physics-based approaches are emerging as powerful enablers of rational conjugation-site selection in ADC development, mitigating the reliance on empirical trial-and-error strategies. Structural dynamics and solvent accessibility analyses [45,46,57] enable rapid prioritization of candidate conjugation sites, which can be further refined through molecular docking [58].

Atomistic or high-resolution coarse-grained simulations can capture antibody conformational heterogeneity, revealing transient or "cryptic" reactive sites inaccessible in crystallographic static models. In this principle underlies the SILCS-Covalent framework, which exploits protein dynamics to identify buried cysteine residues that become transiently solvent-exposed [59]. Residue reactivity can be further quantified through pKa predictions, with MD-based [60] and machine-learning-based methods [61] providing complementary strategies to identify nucleophilic sites with high modification potential. Physics-based modeling has also enabled non-covalent ADC design paradigms. For instance, combined docking and MD simulations have identified stable antibody binding sites for 4-mercaptoethylpyridine, a small-molecule anchor used to recruit linker–payload constructs [58].

**Linker and payload interactions**

Molecular modeling has also revealed that ADCs undergo substantial conformational variability, driven by relative motions of the Fab and Fc domains, and that ADCs structural dynamics are strongly influenced by payload conjugation site, linker and DAR [4,62]. Distinct conformational ensembles have been observed across ADC variants, consistent with the expectation that ADC structure strongly impacts biological activity [63]. Remarkably, most ADCs have exhibited conformational variability comparable to or lower than the unconjugated monoclonal antibody (mAb) [64]. Payload–payload interactions further modulate ADC structural conformations [65–67], a key aspect to take in consideration when designing ADCs with different payloads [68]. In addition, accessibility analyses relevant to cathepsin B–mediated drug release showed that conjugation reduces payload accessibility relative to free payloads, with single-site ADCs generally exhibiting higher accessibility than higher-DAR constructs [62,69]. Together, these findings highlight the potential advantage of using molecular simulations for the rational optimization and prediction of how conjugation site, linker, and DAR reshape the antibody multivalency, dynamic structural ensemble, and overall phase behavior from receptor recognition to successful membrane internalization [38].

**Chemical Manufacturing and Control (CMC) properties**

Chemical and manufacturing strategies are typically addressed as downstream engineering challenges, often optimized after key molecular design decisions—such as conjugation site, linker chemistry, and DAR—have been fixed. However, growing evidence indicates that many CMC-relevant properties, including aggregation propensity, solubility, and excipient sensitivity, are strongly influenced by the underlying intermolecular interactions and conformational ensemble distributions encoded across the sequence and conjugation sites established early in design. This perspective focuses on the biophysical principles underlying antibody and ADC phase behavior at molecular and cellular scales, rather than on formulation and manufacturing optimization strategies previously reviewed [4,70]; nonetheless, computational predictions play a crucial role in de-risking how sequence modifications alter mesoscale phase behavior and impact clinical efficacy. In that sense, CG simulations, such as those using the 12-bead model [38], have been successfully used to provide mechanistic insights into antibody aggregation propensities and viscosity changes at high concentrations [39]. Moreover, CG models have enabled the exploration of how conjugation-induced modifications in antibody surface properties affect colloidal stability and higher-order self-assembly, complementing atomistic predictions of linker–payload behaviour [4]. High-resolution CG force fields, such as MARTINI—or its different subsequent improved reparameterizations can also contribute to characterizing protein−excipient and protein-membrane interactions [35]. By bridging the resolution gap between individual biomolecular interactions and larger mesoscale ensembles, CG simulations enable the systematic mapping of formulation effects and sequence modifications, thereby offering a scalable route to de-risk ADC design prior to experimental validation.

**Limitations of multiscale modeling: antibodies versus ADCs**

The key limitations of multiscale computational modeling manifest differently depending on whether the therapeutic of interest is a naked antibody or an ADC. For antibodies, the primary challenges concern adequate conformational sampling of Fab–Fc interdomain flexibility and the accurate prediction of self-association at high concentrations—a regime where force-field accuracy at protein–protein interfaces remains a critical bottleneck, as small systematic errors in interaction energies can translate into large errors in predicted aggregation propensity. For ADCs, an additional and distinct layer of complexity arises on multiple fronts. First, heterogeneous small-molecule linker–payload constructs must be accurately parameterized, a non-trivial task given that validated force-field parameters for novel payload chemistries—including emerging topoisomerase inhibitors and RNA polymerase inhibitors—are frequently unavailable, necessitating bespoke parameterization efforts that substantially increase computational overhead. Second, DAR-dependent conformational ensemble redistribution must be captured: as the number of attached payloads increases, the global shape and dynamic behaviour of the ADC changes in ways that affect colloidal stability, receptor engagement, and biodistribution, and these changes differ across DAR species within the same preparation. Third, payload–payload interactions—which do not exist in naked antibodies—can drive intermolecular aggregation or sterically shield enzymatic cleavage sites from intracellular proteases, directly impairing drug release and efficacy. Together, these distinctions mean that the proposed multiscale framework is most mature and immediately applicable for antibody-level modeling, while ADC-specific applications require continued methodological investment—particularly in transferable small-molecule parameterization, enhanced sampling of conjugated ensembles, and experimental benchmarking of DAR-resolved structural predictions.

## Concluding remarks and future directions

A central future direction for antibody and ADC development is the systematic translation of biophysical insight—from molecular-level interactions to mechanistic self-assembly and supramolecular organization—into early, decision-level design guidance. Beyond their traditional interpretive role, computational biophysical models are increasingly positioned to inform specific design choices at defined stages of development. Crucially, the nature of these choices differs between naked antibodies and ADCs: antibody design centres on binding affinity, selectivity, and colloidal stability, whereas ADC design demands simultaneous co-optimisation of the linker and payload alongside the antibody scaffold, requiring models that account for conjugation-induced conformational changes, payload–antibody interactions, DAR effects, and intracellular release kinetics. Hydrophobic payloads—such as auristatins, maytansinoids, and camptothecin derivatives—introduce aggregation propensity and non-specific membrane interactions absent in unconjugated antibodies, while linker heterogeneity in stochastic conjugation strategies generates DAR-dependent colloidal behaviour requiring ADC-specific modeling considerations (see Outstanding Questions). While the overarching multiscale framework applies to both modalities, modeling priorities shift substantially: colloidal stability and binding affinity dominate for naked antibodies, whereas payload–antibody coupling, cleavage site accessibility, and DAR-driven phase behaviour require equal or greater attention for ADCs.

At the early discovery stage, ensemble-based analyses generated through MD simulations at different resolutions can guide target and epitope prioritization, as well as affinity tuning, by identifying binding regimes that balance receptor engagement, multivalency, and productive internalization rather than simply maximizing binding affinity. During lead optimization, atomistic and high-resolution models can inform conjugation site selection, linker architecture, and the feasible DAR window, helping identify designs that preserve antigen recognition while limiting unfavourable payload–antibody and payload–payload interactions. At this stage, computational models act as an exclusionary filter, eliminating candidates prone to failure from occluded cleavage sites, excessive conformational restriction, or destabilisation of binding domains through undesired cross-interactions (see Outstanding Questions)

A key advantage of physics-based platforms is their applicability to novel ADC architectures without reliance on prior training data—in contrast to purely machine learning–based approaches, whose performance depends heavily on dataset availability and representativeness [71]. By delivering mechanistic, submolecular-resolution rationale, multiscale MD frameworks can compress design timelines and de-risk translational development. Nevertheless, important limitations remain: the computational cost of large system sizes, force-field accuracy for novel payload chemistries, and challenges in achieving adequate conformational sampling and convergence[72]. Ongoing advances in enhanced-sampling methodologies, GPU-parallelized simulation pipelines[21] , improved parameterization strategies, and hybrid physics–machine-learning approaches[73] are expected to accelerate the transition from discovery to clinic[74] —and may ultimately enable the *de novo* design of antibodies and ADCs with atomic-level precision in both structure and epitope targeting.

**Acknowledgements**
Alberto Ocana's lab is supported by CRIS Cancer Foundation (AOF. C01, AOF.M01), Instituto de Salud Carlos III (PI19/00808), ACEPAIN Foundation, and CIBERONC. J. R. E. acknowledges funding from Emmanuel College, the University of Cambridge, the Ramón y Cajal fellowship (RYC2021-030937-I), the Spanish scientific plan and committee for research reference PID2022-136919NA-C33, and the European Research Council (ERC) under the European Union's Horizon Europe research and innovation program (grant agreement no. 101160499).

**Author contributions**
A.O. and J. R. E. conceived the article and prepared the final version.

**Declaration of interests**
A.O. Consultant fee from NMS. Former consultant of Servier, WWIT and CancerAppy. Former employee of Symphogen. A.O. and J.R.E are both co-founders of C-Therapeutics.

**Figure 1. Multiscale Computational Frameworks for Antibody Structure Prediction and Property Characterization** a) Left: Schematic representation of an antibody molecular structure highlighting the constant (C) and variable (V) domains of the light (L) and heavy (H) chains. Right: Atomistic structural prediction of an antibody obtained using sequence reconstruction with the open-access AlphaFold/ColabFold softwares **(b)** Antibody properties that can be investigated using different computational approaches, including all-atom simulations (orange panel), coarse-grained models (green panel), and SILCS-Biologics hybrid methods (cyan panel). These properties and methodologies are displayed as a function of computational cost, system size and complexity, and achievable accuracy.

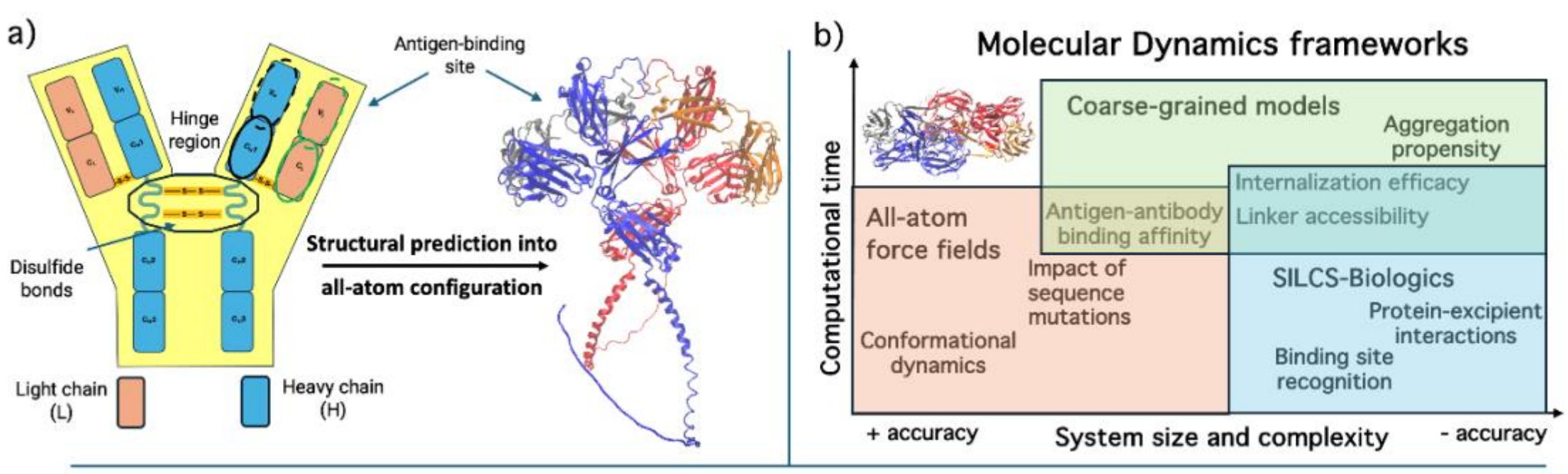

**Figure 2. From All-Atom to Coarse-Grained Simulations and Experimental Validation of Antibody Properties (a)** Workflow illustrating how an all-atom antibody structure can be coarse-grained into a high-resolution coarse-grained model (e.g., MARTINI) to investigate mesoscale processes such as antibody aggregation on membranes or internalization efficiency as a function of antibody surface concentration. Through a back-mapping protocol selected pre-equilibrated antibody–antigen complexes can be converted back to all-atom resolution to quantify binding affinities with higher accuracy than coarse-grained simulations alone. **(b)** Properties which can be evaluated both experimentally and computationally to benchmark modeling predictions.

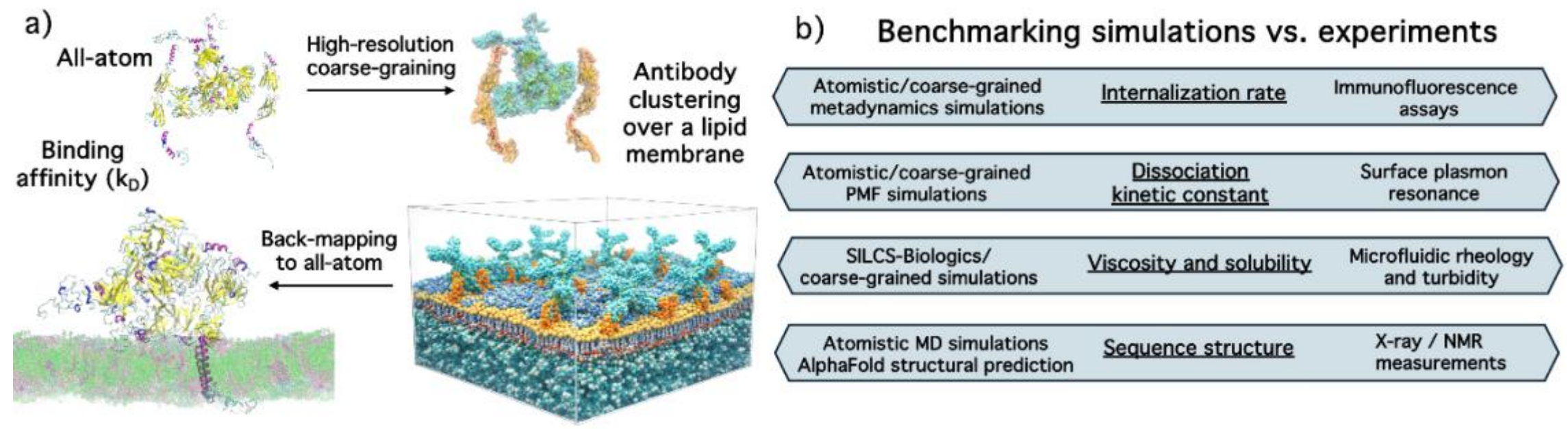